\documentclass[conference]{IEEEtran}
\usepackage{amsmath,amssymb,amsfonts,epsfig}
\usepackage{verbatim}
\usepackage{graphicx}
\usepackage{graphics}

\newcommand{\ber}{\begin{eqnarray}}
\newcommand{\eer}{\end{eqnarray}}

\newtheorem{theorem}{\noindent Theorem}
\newtheorem{lemma}{\noindent Lemma}

\newtheorem{proposition}{\noindent Proposition}

\newcommand{\be}{\begin{equation}}
\newcommand{\ee}{\end{equation}}

\newcommand{\eps}{\epsilon}

\newcommand{\nin}{\noindent}

\begin{document}

\title{Taylor series expansions for the entropy rate of Hidden Markov Processes}

\author{\authorblockN{Or Zuk \quad Eytan Domany}
\authorblockA{Dept. of Physics of Complex Systems\\
Weizmann Inst. of Science\\
Rehovot, 76100, Israel\\
Email: \{or.zuk\}/\{eytan.domany\}@weizmann.ac.il} \and
\authorblockN{Ido Kanter}
\authorblockA{Faculty of Physics\\
Bar-Ilan Univ.\\
Ramat-Gan, 52900, Israel\\
Email: kanter@mail.biu.ac.il} \and
\authorblockN{Michael Aizenman}
\authorblockA{Deptartment of Physics \\
Princeton Univ.\\
Princeton, NJ 08544-0708\\
Email: aizenman@princeton.edu}}

\maketitle

\begin{abstract}
Finding the entropy rate of Hidden Markov Processes is an active
research topic, of both theoretical and practical importance. A
recently used approach is studying the asymptotic behavior of the
entropy rate in various regimes. In this paper we generalize and
prove a previous conjecture relating the entropy rate to entropies
of finite systems. Building on our new theorems, we establish
series expansions for the entropy rate in two different regimes.
We also study the radius of convergence of the two series
expansions.
\end{abstract}

\IEEEpeerreviewmaketitle

\section{Introduction}
Let $\{X_N\}$ be a finite state stationary markov process over the
alphabet $\Sigma = \{1,\ldots,s\}$. Let $\{Y_N\}$ be its noisy
observation (on the same alphabet). Let $M = M_{s \times s} =
\{m_{ij}\}$ be the Markov transition matrix and $R = R_{s \times
s}$ be the emission matrix, i.e. $P(X_{N+1} = j | X_N = i) =
m_{ij}$ and $P(Y_N = j | X_N = i) = r_{ij}$. We assume that the
Markov matrix $M$ is strictly positive ($m_{ij} > 0$), and denote
its stationary distribution by the (column) vector $\pi$
,satisfying $\pi^t M = \pi^t$.

\nin The process $Y$ can be viewed as a noisy observation of $X$,
through a noisy channel. It is known as a {\it Hidden Markov Process
(HMP)}, and is determined by the parameters $M$ and $R$. More
generally, {\it HMPs} have a rich and developed theory, and
enourmous applications in various fields (see
\cite{Merhav,Rabiner}).

\nin An important property of the process $Y$  is its entropy
rate. The Shannon entropy rate of a stochastic process
(\cite{Shannon}) measures the amount of 'uncertainty per-symbol'.
More formally, for $i \leq j$, let $[X]_i^j$ denote the vector
$(X_i,\ldots,X_j)$. Then the entropy rate $\bar{H}(Y)$ is defined
as : \be \bar{H}(Y) = \lim_{N \to \infty} \frac{H([Y]_1^N)}{N}
\label{entropy_rate_def}\ee

\nin Where $H(X) = -\sum_X P(X) \log P(X)$; Here and throughout
the paper we use natural logarithms, so the entropy is measured in
{\it NATS}, and also adopt the convention $0 \log 0 \equiv 0$. We
sometimes omit the realization $x$ of the variable $X$, so $P(X)$
should be understood as $P(X = x)$. The entropy rate can also be
computed via the conditional entropy as: $\bar{H}(Y) = \lim_{N \to
\infty} H(Y_N | [Y]_1^{N-1})$, since for a stationary process the
two limits exist and coincide (\cite{Cover}). The conditional
entropy $H(Y|X)$ (where $X,Y$ are sets of r.v.s.) represents the
average uncertainty of $Y$, assuming that we know $X$, that is
$H(Y|X) = \sum_{x} P(X = x) H(Y | X=x) $. By the chain rule for
entropy, it can also be viewed as a difference of entropies,
$H(Y|X) = H(X,Y) - H(X)$, which will be used later.

\nin There is at present no explicit expression for the entropy
rate of a {\it HMP} (\cite{Merhav,Jacquet}). Few recent works
(\cite{Jacquet, Weissman:02,ZKD1}) have dealt with finding the
asymptotic behavior of $\bar{H}$ in several parameter regimes.
However, they concentrated only on binary alphabet, and proved
rigorously only bounds or at most second (\cite{ZKD1}) order
behavior.

\nin Here we generalize and prove a conjecture posed in \cite{ZKD1},
which justifies (under some mild assumptions) the computation of
$\bar{H}$ as a series expansion in the High Signal-to-Noise-Ratio
('High-SNR') regime. The expansion coefficients were given in
\cite{ZKD1}, for the symmetric binary case. In this case, the
matrices $M$ and $R$ are given by :
$$
M = \begin{pmatrix}
  1-p & p \\
  p & 1-p
\end{pmatrix} \: \:, \: \:
R = \begin{pmatrix}
  1-\eps & \eps \\
  \eps & 1-\eps
\end{pmatrix}
$$

\nin and the process is characterized by the two parameters
$p,\eps$. The High-SNR expansion in this case is an expansion in
$\eps$ around zero.

\nin In section \ref{two_thms_sec}, we present and prove our two
main theorems; Thm. \ref{high_SNR_thm} is a generalization of a
conjecture raised in \cite{ZKD1} which connects the coefficients
of entropies using finite histories to the entropy rate. Proving
it justifies the High-SNR expansion of \cite{ZKD1}. We also give
Thm. \ref{almost_memoryless_thm}, which is the analogous of Thm.
\ref{high_SNR_thm} in a different regime, termed
'Almost-Memoryless' ('A-M').

\nin In section \ref{series_coeff_sec} we use our two new theorems
to compute the first coefficients in the series expansions for the
two regimes. We give the first-order asymptotics for a general
alphabet, as well as higher order coefficients for the symmetric
binary case.

\nin In section \ref{analytic_sec} we estimate the radius of
convergence of our expansions using a finite number of terms, and
compare our results for the two regimes. We end with conclusions and
future directions.

\section{From Finite system entropy to entropy rate}
\label{two_thms_sec} In this section we prove our main results,
namely Thms. \ref{high_SNR_thm} and \ref{almost_memoryless_thm},
which relate the coefficients of the finite bounds $C_N$ to those
of the entropy rate $\bar{H}$ in two different regimes.

\subsection{The High SNR Regime}

\label{high_SNR_sec} \nin This regime was dealt in further details
in \cite{ZKD1,ZKD2}, albeit with no rigorous justification for the
obtained series expansion. In the High-SNR regime the observations
are likely to be equal to the states, or in other words, the
emission matrix $R$ is close to the identity matrix $I$. We
therefore write $R = I + \eps T$, where $\eps > 0$ is a small
constant and $T = \{t_{ij}\}$ is a matrix satisfying $t_{ii} < 0,
\: t_{ij} \geq 0 , \: \forall i \neq j$ and
$\sum_{j=1}^{s}{t_{ij}} = 0$. The entropy rate in this regime can
be given as an expansion in $\eps$ around zero. We state here our
new theorem, connecting the entropy of finite systems to the
entropy rate in this regime.

\begin{theorem}
Let $H_N \equiv H_N(M,T,\eps) = H([Y]_1^N)$  be the entropy of a
finite system of length $N$, and let $C_N = H_N - H_{N-1}$.
Assume\footnote{It is easy to show that the functions $C_N$ are
differentiable to all orders in $\eps$, at $\epsilon =0$. The
assumption which is not proven here is that they are in fact
analytic with a radius of analyticity which is uniform in N, and are
uniformly bounded within some common neighborhood of $\epsilon =0$}
that there is some (complex) neighborhood $B_{\rho}(0) = \{\eps :
|\eps| < \rho \} \subset \mathbb{C}$ of $\eps=0$, in which the
(one-variable) functions $\{ C_N \}, \bar{H}$ are analytic in
$\eps$, with a Taylor expansion given by : \be C_N(M,T,\eps) =
\sum_{k=0}^{\infty} C_N^{(k)} \eps^k, \quad \bar{H}(M,T,\eps) =
\sum_{k=0}^{\infty} C^{(k)} \eps^k \ee

\nin (The coefficients $C_N^{(k)}$ are functions of the parameters
$M$ and $T$. From now on we omit this dependence). Then: \be N \geq
\lceil \frac{k+3}{2} \rceil \Rightarrow C_N^{(k)} = C^{(k)} \ee
\label{high_SNR_thm}
\end{theorem}

\nin The recent result (\cite{Marcus}) on analyticity of $\bar{H}$
is not applicable near $\eps=0$, therefore the analytic domain of
$C_N$,
and, more importantly $\bar{H}$, will be discussed elsewhere. \\

\nin $C_N$ is actually an upperbound (\cite{Cover}) for $\bar{H}$.
The behavior stated in Thm. \ref{high_SNR_thm} was discovered
previously using symbolic computations, but was proven only for $k
\leq 2$ , and only for the symmetric binary case (see \cite{ZKD1}).

\nin Although technically involved , the proof of Thm.
\ref{high_SNR_thm} is based on the following two simple ideas.
First, we distinguish between the noise parameters at different
sites. This is done by considering a more general process $\{Z_N\}$,
where $Z_i$'s emission matrix is $R_i = I + \eps_iT$. The joint
distribution of $[Z]_1^N$ is thus determined by $M$,$T$ and
$[\eps]_{1}^N$. We define the following functions : \be
F_N(M,T,[\eps]_1^{N}) = H([Z]_1^N) - H([Z]_1^{N-1}) \ee

\nin Setting all the $\eps_i$'s equal, reduces us back to the $Y$
process, so in particular $F_N(M,T, (\eps,\ldots,\eps)) =
C_N(\eps)$.

\nin Second, we observe that if a particular $\eps_i$ is set to
zero, the corresponding observation $Z_i$ must equal the state
$X_i$. Thus, conditioning back to the past is 'blocked'. This can
be used to prove the following :

\begin{lemma}

Assume $\eps_j=0$ for some $1 < j < N$. Then :
$$
F_N([\eps]_1^N) = F_{N-j+1}([\eps]_{j+1}^N)
$$

\begin{proof}

\nin $F$ can be written as a sum of conditional entropies :

\be F_N = -\sum_{[Z]_1^N} P([Z]_1^{N-1}) P(Z_N | [Z]_1^{N-1}) \log
P(Z_N | [Z]_1^{N-1}) \label{F_cond_eq}\ee

\nin Where the dependence on $[\eps]_1^N$ and $M,T$ comes through
the probabilities $P(..)$. Since $\eps_j=0$, we must have $X_j =
Z_j$, and therefore (since the $X_i$'s form a Markov chain),
conditioning further to the past is 'blocked', that is : \be \eps_j
= 0 \Rightarrow P(Z_N | [Z]_1^{N-1}) = P(Z_N | [Z]_j^{N-1})
\label{blocking_cond} \ee

\nin (Note that eq. (\ref{blocking_cond}) is true for $j < N$, but
not for $j=N$). Substituting in eq. (\ref{F_cond_eq}) gives :
$$
F_N = -\sum_{[Z]_1^N} P([Z]_1^{N-1}) P(Z_N | [Z]_j^{N-1}) \log
P(Z_N | [Z]_j^{N-1}) =
$$
$$
-\sum_{Z_{j}^N} P([Z]_j^{N-1}) P(Z_N | [Z]_j^{N-1}) \log P(Z_N |
[Z]_j^{N-1})
$$
\be
  = F_{N-j+1}
\ee

\end{proof}

\label{F_cond_lemma}
\end{lemma}

\nin Let $\vec{k} = [k]_1^N$ be a vector with $k_i \in
\{\mathbb{N} \cup 0\}$. Define its 'weight' as $\omega(\vec{k}) =
\sum_{i=1}^N k_i$. Define also : \be F_N^{\vec{k}} \equiv \left.
\frac{\partial^{\omega(\vec{k})} F_N}{\partial
\eps_1^{k_1},\ldots,\partial \eps_N^{k_N}}
 \right|_{\vec{\eps} = 0}
\ee

\nin The next lemma shows that adding zeros to the left of
$\vec{k}$ leaves $F_N^{\vec{k}}$ unchanged :
\begin{lemma}
Let $\vec{k} = [k]_1^N$ with $k_1 \leq 1$. Denote $\vec{k}^{(r)}$
the concatenation of $\vec{k}$ with $r$ zeros : $\vec{k}^{(r)} =
(\underbrace{0,\ldots,0}_r,k_1,\ldots,k_N)$. Then :
$$
 F_N^{\vec{k}} = F_{r+N}^{\vec{k}^{(r)}} \quad,  \forall r \in \mathbb{N}
$$

\begin{proof}
\nin Assume first $k_1 = 0$. Using lemma \ref{F_cond_lemma}, we get
:
$$
F_{N+r}^{\vec{k}^{(r)}}([\eps]_1^{N+r}) = \left.
\frac{\partial^{\omega(\vec{k}^{(r)})}
F_{r+N}([\eps]_1^{N+r})}{\partial \eps_{r+2}^{k_2},\ldots,\partial
\eps_{r+N}^{k_N}}
 \right|_{\vec{\eps} = 0} =
$$
\be
 \left.
\frac{\partial^{\omega(\vec{k})}
F_{N}([\eps]_{r+1}^{N+r})}{\partial
\eps_{r+2}^{k_2},\ldots,\partial \eps_{r+N}^{k_N}}
 \right|_{\vec{\eps} = 0} = F_N^{\vec{k}}([\eps]_{r+1}^{r+N})
 \label{k_is_zero_eq}
\ee

\nin The case $k_1 = 1$ is reduced back to the case $k_1=0$ by
taking the derivative. We denote by ${[Z]_1^N}^{(j \to r)}$ the
vector which is equal to $[Z]_1^N$ in all coordinates except on
coordinate $j$, where $Z_j = r$. Using eq. (\ref{k_is_zero_eq}),
we get :
$$
F_{N+1}^{\vec{k}^{(1)}}([\eps]_1^{N+1}) = \left.
\frac{\partial^{\omega(\vec{k})-1}}{\partial \eps_3^{k_2} \dots
\partial \eps_{N+1}^{k_N}} \left[ \left. \frac{\partial
F_{N+1}}{\partial \eps_2} \right|_{\eps_2 = 0} \right]
\right|_{\vec{\eps} = 0} =
$$
$$
\frac{\partial^{\omega(\vec{k})-1}}{\partial \eps_3^{k_2} \dots
\partial \eps_{N+1}^{k_N}} \Biggl\{ - \sum_{r=1}^{s} t_{X_i r}
\sum_{[Z]_1^{N+1}}
$$
$$
  \left[ P({[Z]_1^{N+1}}^{(2 \to r)})
\log P(Z_{N+1} | [Z]_1^{N}) - \right.
$$
$$
\left. \left.  \left. P(Z_{N+1} | [Z]_1^{N}) P({[Z]_1^{N}}^{(2 \to
r)}) \right]  \right|_{\eps_2 = 0} \Biggr\}
\right|_{[\eps]_1^{N+1} = 0} =
$$
$$
\frac{\partial^{\omega(\vec{k})-1}}{\partial \eps_2^{k_2} \dots
\partial \eps_{N}^{k_N}} \Biggl\{ - \sum_{r=1}^{s} t_{X_i r}
\sum_{[Z]_1^{N}}
$$
$$
  \left[ P({[Z]_1^{N}}^{(1 \to r)})
\log P(Z_{N} | [Z]_1^{N-1}) - \right.
$$
\be \left. \left.  \left. P(Z_{N} | [Z]_1^{N-1}) P({[Z]_1^{N}}^{(1
\to r)}) \right]  \right|_{\eps_1 = 0} \Biggr\}
\right|_{[\eps]_1^{N} = 0} = F_{N}^{\vec{k}}([\eps]_1^{N})\ee

\end{proof}

\label{zero_tail_lemma}
\end{lemma}

\nin $C_N^{(k)}$ is obtained by summing $F_N^{\vec{k}}$ on all
$\vec{k}$'s with weight $k$ : \be C_N^{(k)} =
\sum_{\vec{k},\omega(\vec{k})=k} F_N^{\vec{k}} \ee

\nin We now show that one does not need to sum on all such
$\vec{k}$'s, as many of them give zero contribution :

\begin{lemma}
Let $\vec{k} = (k_1,\ldots,k_N)$. If $\exists i < j < N$, with
$k_i \geq 1, k_j \leq 1$, then $F_N^{\vec{k}} = 0$.

\begin{proof}

\nin Assume first $k_j = 0$. Using lemma \ref{F_cond_lemma} we get

$$
F_N^{\vec{k}} \equiv \left. \frac{\partial^{\omega(\vec{k})}
F_N(\vec{\eps})}{\partial \eps_1^{k_1},\ldots,\partial
\eps_N^{k_N}}
 \right|_{\vec{\eps} = 0} = \left. \frac{\partial^{\omega(\vec{k})}
F_{N-j+1}([\eps]_j^N)}{\partial \eps_1^{k_1},\ldots,\partial
\eps_N^{k_N}}
 \right|_{\vec{\eps} = 0} =
$$
\be \frac{\partial^{\omega(\vec{k})-1}}{\partial
\eps_1^{k_1},\ldots,
\partial \eps_i^{k_i-1},\ldots, \partial \eps_N^{k_N}} \left[\left.
\frac{\partial F_{N-j+1}([\eps]_j^N)}{\partial \eps_i} \right]
\right|_{\vec{\eps} = 0} = 0 \ee

\nin The case $k_j = 1$ is more difficult, but follows the same
principles. Write the probability of $Z$ :
$$
 P([Z]_1^N)  = \sum_{[X]_1^N} P([X]_1^N) P([Z]_1^N |
[X]_1^N) =
$$
\be \sum_{[X]_1^N} P([X]_1^N) \prod_{i=1}^N (\delta_{X_i Z_i} +
\eps_i t_{X_i Z_i})\ee

\nin where $\delta_{ij}$ is Kronecker delta. Write now the
derivative with respect to $\eps_j$:

$$
\left. \frac{\partial P([Z]_1^N)}{\partial \eps_j} \right|_{\eps_j
= 0} =
$$
$$
\left. \sum_{[X]_1^N} \left[ P([X]_1^N) t_{X_j Z_j} \prod_{i \neq
j} (\delta_{X_i Z_i} + \eps_i t_{X_i Z_i}) \right] \right|_{\eps_j
= 0} =
$$
\be \left. \left\{ \sum_{r=1}^{s} t_{X_i r} P({[Z]_1^N}^{(j \to
r)}) \right\} \right|_{\eps_j = 0} \ee

\nin Using Bayes' rule $P(Z_N | [Z]_1^{N-1}) =
\frac{P([Z]_1^N)}{P([Z]_1^{N-1})}$, we get :

$$
\left. \frac{\partial P(Z_N | [Z]_1^{N-1})}{\partial \eps_j}
\right|_{\eps_j = 0} =
$$
$$
\frac{1}{P([Z]_1^{N-1})} \sum_{r=1}^{s} t_{X_i r} \left[
P({[Z]_1^N}^{(j \to r)}) - \right.
$$
\be \left. \left. P(Z_N | [Z]_1^{N-1} ) P({[Z]_1^{N-1}}^{(j \to
r)}) \right] \right|_{\eps_j = 0} \ee

\nin This gives :
$$
\left. \frac{\partial [P([Z]_1^N) \log P(Z_N |
[Z]_1^{N-1})]}{\partial \eps_j} \right|_{\eps_j = 0} =
$$
$$
\sum_{r=1}^{s} t_{X_i r} \left\{ P({[Z]_1^N}^{(j \to r)}) \log
P(Z_N | [Z]_1^{N-1}) + \right.
$$
\be \left. \left. P({[Z]_1^N}^{(j \to r)})  - P(Z_N | [Z]_1^{N-1})
P({[Z]_1^{N-1}}^{(j \to r)})\right\} \right|_{\eps_j = 0} \ee

And therefore :
$$
\left. \frac{\partial F_N}{\partial \eps_j} \right|_{\eps_j = 0} =
$$
$$
 -\sum_{r=1}^{s} t_{X_i r} \Biggl\{ \sum_{[Z]_1^N} \left[
P({[Z]_1^N}^{(j \to r)}) \log P(Z_N | [Z]_1^{N-1}) - \right.
$$
$$
 \left. \left. P(Z_N | [Z]_1^{N-1}) P({[Z]_1^{N-1}}^{(j \to
r)}) \right] \Biggr\} \right|_{\eps_j = 0} =
$$
$$
\Biggl\{ -\sum_{r=1}^{s} t_{X_i r} \sum_{[Z]_j^N} \left[
P({[Z]_j^N}^{(1 \to r)}) \log P(Z_N | [Z]_j^{N-1}) - \right.
$$
\be \left.  \left. P(Z_N | [Z]_j^{N-1}) P({[Z]_j^{N-1}}^{(1 \to
r)}) \right] \Biggr\} \right|_{\eps_1 = 0} \label{F_N_deriv_eq}
\ee

\nin Where the latter equality comes from using  eq.
(\ref{blocking_cond}), which 'blocks' the dependence backwards.
Eq. \ref{F_N_deriv_eq} shows that $\left. \frac{\partial
F_N}{\partial \eps_j} \right|_{\eps_j = 0}$ does not depend on
$\eps_i$ for $i < j$, therefore $\frac{\partial^{k_i+1}
F_N}{\partial \eps_i^{k_i}
\partial \eps_j} = 0$ and $F_N^{\vec{k}} = 0$.

\end{proof}
\label{No_hole_strong_lemma}
\end{lemma}

\nin We are now ready to prove Thm. \ref{high_SNR_thm}, which
follows directly from lemmas \ref{zero_tail_lemma} and
\ref{No_hole_strong_lemma} :

\begin{proof}

\nin Let  $\vec{k} = [k]_1^N$ with $\omega(\vec{k}) = k$. Define
its 'length' (from right, considering only entries larger than
one) as $l(\vec{k}) = N+1-\min_{k_i > 1} \{i \}$. It easily
follows from lemma \ref{No_hole_strong_lemma} that if
$F_N^{\vec{k}} \neq 0$, we must have $l(\vec{k}) \leq \lceil
\frac{k+3}{2} \rceil - 1$. Therefore, according to lemma
\ref{zero_tail_lemma} we have :

\be F_N^{\vec{k}} = F_{\lceil \frac{k+3}{2} \rceil}^{(k_{N-\lceil
\frac{k+3}{2} \rceil+1},\ldots,{k_N})} \ee

\nin for all $\vec{k}$'s in the sum. Summing on all
$F_N^{\vec{k}}$ with the same 'weight', we get $C_N^{(k)} =
C_{\lceil \frac{k+3}{2} \rceil}^{(k)}, \quad \forall N
> \lceil \frac{k+3}{2} \rceil$. From the analyticity
of $C_N$ and $\bar{H}$ around $\eps=0$, one can show by induction
that $\lim_{N \to \infty} C_N^{(k)} = C^{(k)}$, therefore we must
have $C_N^{(k)} =
C^{(k)}, \quad \forall N \geq \lceil \frac{k+3}{2} \rceil $. \\

\end{proof}
\label{strong_thm}

\subsection{The Almost Memoryless Regime}
\label{almost_memoryless_sec}

\nin In the A-M regime, the Markov transition matrix is close to
uniform. Thus, throughout this section, we assume that $M$ is
given by $M = U + \delta T$, such that $U$ is a constant (uniform)
matrix, $u_{ij} = s^{-1}$, $\delta > 0$ is a small constant and
$T$ satisfies $\sum_{j=1}^{s}{t_{ij}} = 0$. Thus the process is
entirely characterized by the set of parameters $(R,T,\delta)$,
where $R$ again denotes the emission matrix.

\nin Interestingly, similarly to the High-SNR regime, the
conditional entropy given a finite history gives the correct
entropy rate up to a certain order which depends on the finite
history taken. In the A-M regime we can also prove analyticity of
$\{C_N\}$ and $\bar{H}$ in $\delta$ near $\delta=0$. This is
stated as :

\begin{theorem}
Let $H_N \equiv H_N(R,T,\delta) = H([Y]_1^N)$  be the entropy of a
finite system of length $N$, and let $C_N = H_N - H_{N-1}$. Then :
\begin{enumerate}
\item
There is some (complex) neighborhood $B_{\rho}(0) = \{\delta :
|\delta| < \rho \} \subset \mathbb{C}$ of $\delta=0$, in which the
(one-variable) functions $\{ C_N \}, \bar{H}$ are analytic in
$\delta$, with a Taylor expansion denoted by : \be C_N(M,T,\delta) =
\sum_{k=0}^{\infty} C_N^{(k)} \delta^k, \quad \bar{H}(M,T,\eps) =
\sum_{k=0}^{\infty} C^{(k)} \delta^k \ee (The coefficients
$C_N^{(k)}$ are functions of the parameters $M$ and $T$.)

\item
With the above notations : \be N \geq \lceil \frac{k+3}{2} \rceil
\Rightarrow C_N^{(k)} = C^{(k)} \ee
\end{enumerate}

\label{almost_memoryless_thm}
\end{theorem}

\begin{proof}
\begin{enumerate}
\item
The proof of analyticity relies on the recent result, namely Thm.
1.1 in \cite{Marcus}. In order to use this result, we need to
present the {\it HMP} $Y$ in the following way : We introduce the
new alphabet $\Gamma \subset \Sigma \times \Sigma $ defined by :
$$
\Gamma = \Big\{ w = (w_x,w_y) : w_x,w_y \in \Sigma, r_{w_x w_y}
> 0 \Big\}
$$

We also introduce the function $\Phi : \Gamma \to \Sigma$, defined
by $\Phi(w) \equiv \Phi(w_x,w_y) = w_y$. Let $w,v \in \Gamma$ with
$w = (w_x, w_y), v = (v_x,v_y)$. One can look at the new Markov
process $W = (X,Y)$, defined on $\Gamma$ by the transition matrix
$\Delta_{|\Gamma| \times |\Gamma|}$, which is given by $\Delta_{wv}
\equiv P(W_{N+1} = v | W_N = w) = m_{w_x v_x} r_{v_x v_y}$. Then the
process $Y$ can be defined as $Y_N = \Phi(W_N)$. Using the above
representation, clearly $\Delta$ is analytically parameterized by
$\delta$. Moreover, there is some (real) neighborhood $B_{\rho'}(0)
\subset \mathbb{R}$ in which all of $\Delta$'s entries are positive.
Therefore, Thm. 1.1 from \cite{Marcus} applies here, and according
to its proof, $\{C_N\}$ and $\bar{H}$ are analytic (as functions of
$\delta$) in some complex neighborhood $B_{\rho}(0) \subset
\mathbb{C}$ of zero.

\item
The proof of part 2 is very similar to that of Thm.
\ref{high_SNR_thm}. Distinguishing between the sites by setting
$M_i = U + \delta_i T$ in site $i$, we notice that if one sets
$\delta_i = 0$ for some $i$, then $M_i$ becomes uniform, and thus
knowing $Z_i$ 'blocks' the dependence of $Z_N$ on previous $Z_j$'s
($\forall j < i$). The rest of the proof continues in an analogous
way to the proof of Thm. \ref{high_SNR_thm} (including the three
lemmas therein), and its details are thus omitted here.
\end{enumerate}
\end{proof}

\section{Computation of the series coefficients}
\label{series_coeff_sec}

An immediate application of Thms. \ref{high_SNR_thm} and
\ref{almost_memoryless_thm} is the computation of the first terms
in the series expansion for $\bar{H}$ (assuming its existance), by
simply computing these terms for $C_N$ for $N$ large enough. In
this section we compute, for both regimes, the first order for the
general alphabet case, and also give few higher order terms for
the simple symmetric binary case. Our method for computing
$C^{(k)}$ is straightforward. We compute $C_N^{(k)}$ for $N =
\lceil \frac{k+3}{2} \rceil$ by simply enumerating all sequences
$[Y]_1^N$, computing the $k$-th coefficient in $P([Y]_1^N) \log
P([Y]_1^N)$ for each one, and summing their contribution. This
computation is, however, exponential in $k$, and thus raises the
challenge of designing more efficient algorithms, in order to
compute further orders and for larger alphabets.

\nin Before giving the calculated coefficients, we need some new
notations. For a vector $\alpha$, $diag(\alpha)$ denotes the square
matrix with $\alpha$'s elements on the diagonal. We use Matlab-like
notation to denote element-by-element operations on matrices. Thus,
for matrices $A$ and $B$, $log A$ is a matrix whose elements are
$\{\log a_{ij} \}$, and $[A .* B]$ is a matrix whose elements are
$\{a_{ij} b_{ij} \}$. $\xi$ denotes the (column) vector of $N$ ones.

\subsection{The High-SNR expansion}
According to Thm. \ref{high_SNR_thm}, computing $C_2$ enables us to
extract $\bar{H}^{(k)}$. This is used to show the following :

\begin{proposition}
Let $R = I + \eps T$. Assume that the entropy rate $\bar{H}$ is
analytic in some neighborhood of $\delta = 0$. Then $\bar{H}$ satisfies : 
$$
\bar{H} = -\pi^t [M .* \log M] \xi + \xi^t \Bigl\{ diag(\log (\pi))
T^t diag(\pi) M -
$$
\be [diag(\pi) M T + T^t diag(\pi) M] .* [\log(diag(\pi) M)] \Bigr\}
\xi \eps + O(\eps^2) \label{general_1st_order_final} \ee

\begin{proof}
Noting that according to Thm. \ref{high_SNR_thm}, $\bar{H} = C_2 +
O(\eps^2)$, we first compute (exactly) $C_2$, and then expand it
by substituting $R = I + \eps T$. Write $C_2$ as :
$$
C_2 = H(Y_N | Y_{N-1}) =
$$
\be -\sum_{i,j} P(Y_N = j, Y_{N-1} = i) \log \frac{P(Y_N = j,
Y_{N-1} = i)}{P(Y_{N-1}=i)} \label{C_2_eq}\ee

\nin We can express the above probabilities as :
$$
P(Y_{N-1}=i) = [\pi R]_i
$$
\be P(Y_N = j, Y_{N-1} = i) = [R^t diag(\pi) M R]_{ij} \equiv F_{ij}
\label{prob_ys_eqs}\ee

\nin Substituting eq. (\ref{prob_ys_eqs}) in eq. (\ref{C_2_eq}),
and writing in matrix form, we get : \be C_2 = \Bigl\{ [log(\pi
R)] F - \xi^T [F .* log F] \Bigr\} \xi \label{general_alphabet_
hmm_entropy} \ee

\nin Substituting $R = I + \eps T$ gives :
$$
F = diag(\pi) M + [diag(\pi) M T + T^t diag(\pi) M] \eps +
O(\eps^2),
$$
$$
F .* \log F = [diag(\pi) M] .* \log(diag(\pi) M) +
$$
$$
\Bigl\{[diag(\pi) M T + T^t diag(\pi) M] .* [I + \log(diag(\pi)
M)]\Bigr\} \eps +
$$
\be
O(\eps^2) \label{F_log_F_first_order} \ee

\nin Substituting these in eq. (\ref{general_alphabet_ hmm_entropy})
gives, after simplification, the result
(\ref{general_1st_order_final}).

\end{proof}
\label{high_SNR_first_order}
\end{proposition}

\nin We note that prop. \ref{high_SNR_first_order} above is a
generalization of the result obtained by \cite{Jacquet} for a binary
alphabet.

\nin Turning now into the symmetric binary case, the first eleven
orders of the series expansion were given in \cite{ZKD1}, but only
the first two were proved to be correct. Thm. \ref{high_SNR_thm}
proves the correctness of the entire expansion from \cite{ZKD1}
(under the analyticity assumption on $\bar{H}$), which is not
repeated here.

\subsection{The almost memoryless expansion}

\nin By Thm. \ref{almost_memoryless_thm}, one can expand the
entropy rate around $M=U$ by simply computing the coefficients
$C_N^{(k)}$ for $N$ large enough. For example, by computing $C_2$
we have established, in analogous to prop.
\ref{high_SNR_first_order}, the first order :

\begin{proposition}
Let $M = U + \delta T$.  Then $\bar{H}$ satisfies :
$$
\bar{H} = \log s - s^{-1} \xi^t R [\log (R^t \xi)] -
$$
\be \xi^t \Big[(s^{-1} R^t T R) .* log(s^{-1} R^t U R) \Big] \xi
\delta + O(\delta^2) \label{prop_weak_memory} \ee

\begin{proof}
Since $\bar{H} = C_2 + O(\delta^2)$, we expand $C_2$ (as given in
eq. (\ref{general_alphabet_ hmm_entropy})) in $\delta$. $M$ is
simply replaced by $U + \delta T$. Dealing with $\pi$ is more
problematic. Note that the stationary distribution of $U$ is
$s^{-1} \xi$. We write $\pi = s^{-1} \xi + \delta \psi +
O(\delta^2)$, and solve : \be (s^{-1} \xi + \delta \psi) (U +
\delta T) = (s^{-1} \xi + \delta \psi) + O(\delta^2) \ee

\nin It follows that $\psi$ should satisfy $\psi (I - U) = \xi T$,
where $I$ is the identity matrix. We cannot invert $I-U$ since it
is of rank $s-1$. The extra equation needed for determining $\psi$
uniquely comes from the requirement $\sum_{i=1}^s \psi_i = 0$.
Substituting $M =U + \delta T$ and $\pi = s^{-1} \xi + \psi \delta
+ O(\delta^2)$ in eq. (\ref{general_alphabet_ hmm_entropy}), one
gets :
$$
C_2 = \Big\{log(s^{-1} \xi R) s^{-1} R^t U R -
$$
$$
\xi^t [(s^{-1} R^t U R) .* log (s^{-1} R^t U R)]  \Big\} \xi +
$$
$$
 \bigg\{\log (s^{-1} \xi R) R^t [s^{-1} diag(\xi) T + diag(\psi) U] R -
$$
$$
\xi^t \Big[ \Big(R^t (s^{-1} diag(\xi) T + diag(\psi) U) R \Big)
.*
$$
\be \Big( s U + \log(s^{-1} R^t U R)\Big) \Big] \bigg \} \xi
\delta + O(\delta^2) \label{almost_memoryless_inter_eq}\ee

\nin After further simplification, most terms in eq.
\ref{almost_memoryless_inter_eq} cancel out, and we are left with
the result (\ref{prop_weak_memory}).

\end{proof}
\label{almost_memoryless_first_order}
\end{proposition}

\nin In \cite{Weissman:01} it was shown that the first order term
vanishes for the symmetric binary case, which is consistent with
eq. \ref{prop_weak_memory}. Our result holds for general alphabets
and process parameters. Looking at the symmetric binary case might
be misleading here, since by doing so one fails to see the linear
behavior in $\delta$ for the general case.

\nin We have computed higher orders for the symmetric binary case
by expanding $C_N$ for $N=8$, which gives us $C^{(k)}$ for $k \leq
13$. In this case the expansion is in the parameter $\delta =
\frac{1}{2}-p$, and gives (for better readability the dependency
on $\eps$ is represented here via $\mu = 1-2\eps$) :
$$
\bar{H} = \log(2) - \mu^4 \bigg[2 \delta^2 + \frac{4}{3} (7 \mu^4-12
\mu^2+6) \delta^4+
$$
$$
\frac{32}{15}(46 \mu^8-120 \mu^6+120 \mu^4-60 \mu^2+15) \delta^6 +
$$
$$
\frac{32}{21}(1137 \mu^{12}-4088\mu^{10}+5964 \mu^8-4536
\mu^6+1946 \mu^4-
$$
$$
504 \mu^2+84) \delta^8 + \frac{512}{45} (3346 \mu^{16}-15120
\mu^{14}+28800 \mu^{12}-
$$
$$
30120 \mu^{10}+18990 \mu^8 - 7560 \mu^6+1980 \mu^4-360 \mu^2+45)
\delta^{10} +
$$
$$
\frac{1024}{165} (159230 \mu^{20}-874632 \mu^{18}+ 2091100
\mu^{16}-2857360 \mu^{14}+
$$
$$
2465100 \mu^{12}-1400960 \mu^{10}+532312 \mu^8-135960 \mu^6+
$$
\be 24145 \mu^4- 3300 \mu^2+330) \delta^{12} \bigg] + O(\delta^{14})
; \label{series_almost_memoryless} \ee

\nin The above expansion generalizes a result from
\cite{Weissman:01}, who proved $\bar{H} = \log(2) - 2 \mu^4 \delta^2
+ o(\delta^2)$. Note that for the first few coefficients, all odd
powers of $\delta$ vanish, and the coefficients are all polynomials
of $\mu^2$, which makes this series simpler than the one obtained in
the High-SNR regime (\cite{ZKD1}).

\section{Radius of Convergence}

\label{analytic_sec} The usefulness of a series expansion such as
the ones derived in eq. (\ref{series_almost_memoryless}) and in
\cite{ZKD1} for practical purposes, highly depends on the radius
of convergence. Determining the radius is a difficult problem, as
it relates to the domain of analyticity of $\bar{H}$. In Thm.
\ref{almost_memoryless_thm} we proved that the radius for the A-M
expansion is positive.

\begin{figure}
\centerline{ \psfig{figure=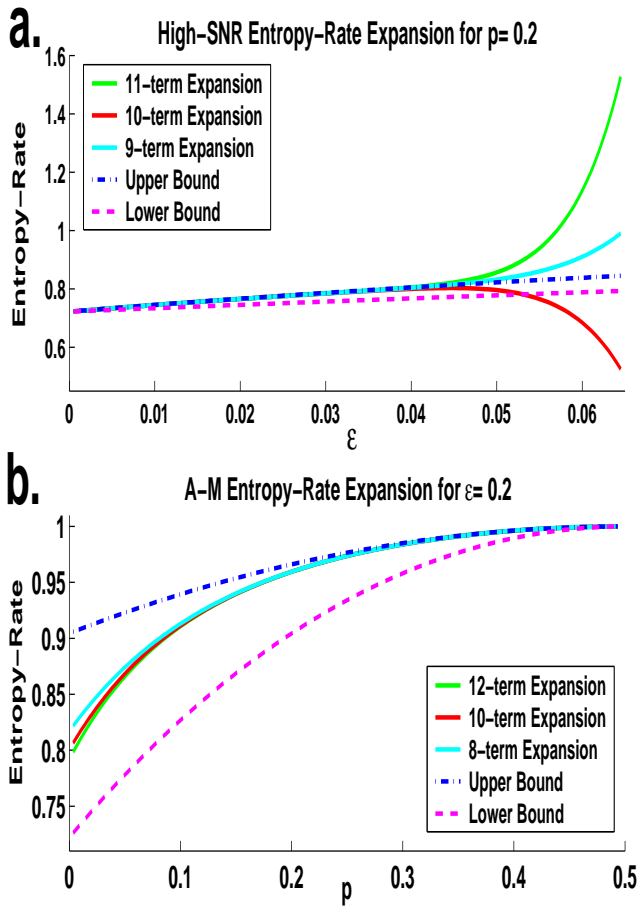,height=12.5cm,width=8.5cm} }  
\caption{Approximations for $\bar{H}$ using first few terms in its
series expansion. a. The High-SNR expansions using $9,10$ and $11$
terms for $p=0.2$ deviate from the bounds for large values of
$\eps$. The first few terms of the expansion have alternating signs,
therefore the direction of the deviation is determined by the parity
of the number of terms taken. b. The A-M expansions using $8,10$ and
$12$ terms for $\eps=0.2$ remain within the bounds for any value of
$p$. \label{TwoExpansionsComparison_fig}} \vspace{1cm}
\end{figure}

\begin{figure}
\centerline{\psfig{figure=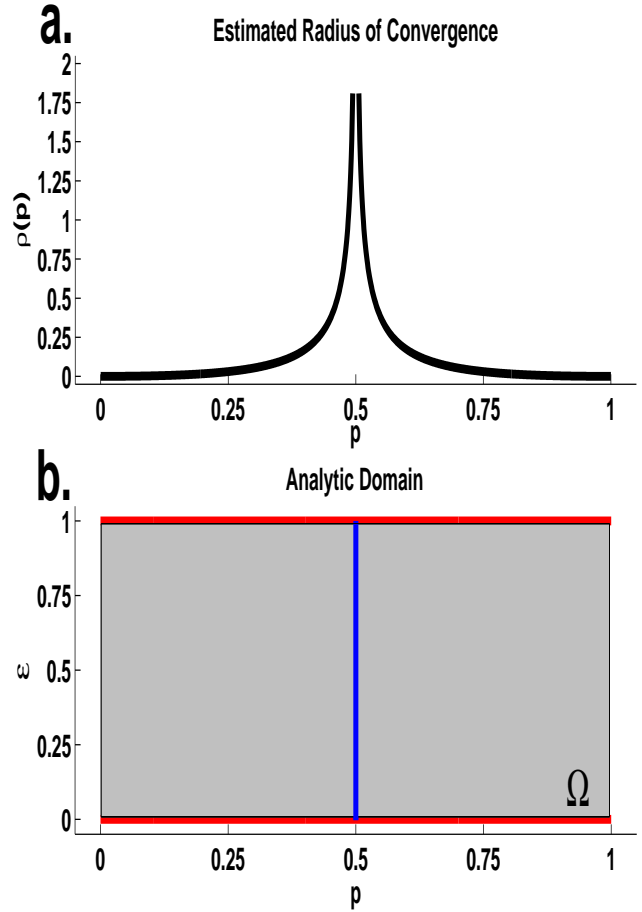,height=12.5cm,width=8.5cm}
} \caption{a. The estimated radius of convergence $\rho(p)$ for
the High-SNR expansion as a function of $p$. b. The domain
$\Omega$ (shaded gray area) in the $\mathbb{R}^2$ plane for which
it is known \cite{Marcus} that $\bar{H}$ is real analytic in
$(p,\eps)$. The A-M expansion is near the vertical line
$p=\frac{1}{2}$. The High-SNR expansion is near the horizonal
boundaries at $\eps=0$ and $\eps=1$. \label{AnalyticDomain_fig}}
\vspace{1cm}
\end{figure}

\nin For the High-SNR case, we gave a numerical estimation of the
radius of convergence $\rho(p)$ as a function of $p$
(\cite{ZKD2}), based on the first few known terms. When one
applies the same procedure to the coefficients of the A-M
expansion, the numerical values of the estimated radius are much
higher. The difference is demonstrated in fig.
\ref{TwoExpansionsComparison_fig}. In this figure, the (finite)
series expansions with up to twelve'th order is compared to two
known bounds on $\bar{H}$ from \cite{Cover}. The upper bound is
simply $C_N = H(Y_N | [Y]_1^{N-1})$ and the lower bound is $c_N
\equiv H(Y_N | X_1, [Y]_1^{N-1})$, for $N=2$. As can be seen from
the figure, for the High-SNR case at $p=0.2$, the finite-order
expansions are not within the bounds for large values of $\eps$.
For the A-M case, for $\eps=0.2$, the finite-order expansions
remain within the bounds for any $0<p<\frac{1}{2}$. \\
\nin The estimated radius $\rho(p)$ for the High-SNR expansion, is
plotted as a function of $p$ in fig. \ref{AnalyticDomain_fig}.a. In
our context, the result of \cite{Marcus} proves that
$\bar{H}(p,\eps)$ is real analytic in the domain $\Omega \subset
\mathbb{R}^2, \quad \Omega = \{(p,\eps): 0<p,\eps<1 \}$ (it is not
known whether $\Omega$ is maximal with that respect). This domain is
shown in fig. \ref{AnalyticDomain_fig}.b. For any $0<\eps<1$, the
A-M expansion is near the point $(\eps, \frac{1}{2})$ which is an
interior point of $\Omega$. The High-SNR expansion is near some
point $(p,0)$, which lies on the boundary of $\Omega$.

\section{Conclusion}
We presented a generalization and proof of the conjecture introduced
in \cite{ZKD1}, relating the expansion coefficients of finite system
entropies to those of the entropy rate for {\it HMPs}. Our new
theorems shed light on the connection between finite and infinite
chains, as well as give a practical and straightforward way to
compute the entropy rate as a series expansion up to an arbitrary
power.

\nin The surprising 'settling' of the expansion coefficients
$C_N^{(k)} = C^{(k)}$ for $N \geq \lceil \frac{k+3}{2} \rceil$,
hold for the entropy. For other functions involving only
conditional probabilities (e.g. relative entropy between two {\it
HMPs}) a weaker result holds: the coefficients 'settle' for $N
\geq k$. We note that this is still a highly non-trivial result,
as it is known that for other regimes (e.g. 'rare-transitions'
\cite{Weissman:03}), a finite chain of any length does not give
the correct asymptotic behavior even to the first order. We also
estimated the radius of convergence for the expansion in the two
regimes, 'High-SNR' and 'A-M', and demonstrated their
quantitatively different behavior. Further research in this
direction, which closely relates to the domain of analyticity of
the entropy rate, is still required.

\section*{Acknowledgment}
M.A. is grateful for the hospitality shown him at the Weizmann
Institute, where his work was supported by the Einstein Center for
Theoretical Physics and the Minerva Center for Nonlinear Physics.
The work of I.K. at the Weizmann Institute was supported by the
Einstein Center for Theoretical Physics. E.D. and O.Z. were
partially supported by the Minerva Foundation and by the European
Community's Human Potential Programme under contract
HPRN-CT-2002-00319, STIPCO.



%

\end{document}